\documentclass[aps,pre,twocolumn]{revtex4}
\usepackage{epsfig}
\usepackage[dvips,colorlinks,linkcolor=blue,citecolor=blue,urlcolor=blue]{hyperref}
\RequirePackage{times}
\usepackage{amssymb}

\begin{document}

\title{Molecular gyroscopes and biological effects of weak ELF magnetic
fields}

\author{V.N. Binhi} \email{Binhi@biomagneti.com}
\homepage{http://www.biomagneti.com} \affiliation{General Physics Institute
Russian Academy of Sciences, 38 Vavilova St., 119991 Moscow, Russia}

\author{A.V. Savin} \email{asavin@center.chph.ras.ru} \affiliation{Institute
for Physics and Technology, 13/7 Prechistenka St., 119034 Moscow, Russia}

%\date{\today}

\begin{abstract} % insert abstract here
Extremely-low-frequency magnetic fields are known to affect biological
systems. In many cases, biological effects display `windows' in biologically
effective parameters of the magnetic fields: most dramatic is the fact that
relatively intense magnetic fields sometimes do not cause appreciable effect,
while smaller fields of the order of 10--100\,$\mu$T do. Linear resonant
physical processes do not explain frequency windows in this case. Amplitude
window phenomena suggest a nonlinear physical mechanism. Such a nonlinear
mechanism has been proposed recently to explain those `windows'. It considers
quantum-interference effects on protein-bound substrate ions. Magnetic fields
cause an interference of ion quantum states and change the probability of
ion-protein dissociation. This ion-interference mechanism predicts specific
magnetic-field frequency and amplitude windows within which biological
effects occur. It agrees with a lot of experiments. However, according to the
mechanism, the lifetime $\Gamma^{-1}$ of ion quantum states within a protein cavity
should be of unrealistic value, more than 0.01 s for frequency band 10--100
Hz. In this paper, a biophysical mechanism has been proposed that (i) retains
the attractive features of the ion interference mechanism, i.e., predicts
physical characteristics that might be experimentally examined and (ii) uses
the principles of gyroscopic motion and removes the necessity to postulate
large lifetimes. The mechanism considers dynamics of the density matrix of
the molecular groups, which are attached to the walls of protein cavities by
two covalent bonds, i.e., molecular gyroscopes. Numerical computations have
shown almost free rotations of the molecular gyros. The relaxation time due
to van der Waals forces was about 0.01 s for the cavity size of 28
angstr\"{o}ms.
\end{abstract}

%\pacs{87.50.M, 87.15, 82.30.F}
%PACS 87.50.M -- Biological effects of magnetic fields \\
%PACS 87.15 -- Molecular biophysics \\
%PACS 82.30.F -- Ion molecule reactions

%\keywords{magnetobiology, quantum mechanics, ion binding, molecular interference}

\maketitle

\section{Introduction} Weak static and extremely-low-frequency (ELF) magnetic
fields (MFs) can affect living things: cells, tissues, physiological systems,
and whole organisms \cite{blank95,goodman95,bersani99}. In many cases
biological effects of weak MF feature resonance-like multipeak behavior.
Multipeak responses or magnetobiological spectra may appear with varying the
frequency or amplitude of AC MF \cite{adey93} and the magnitude of DC MF
\cite{belyaev94c}. Usually, the term `windows' is used for the peaks of the
spectra.

\begin{figure*}[th]
\centering\epsfig{file=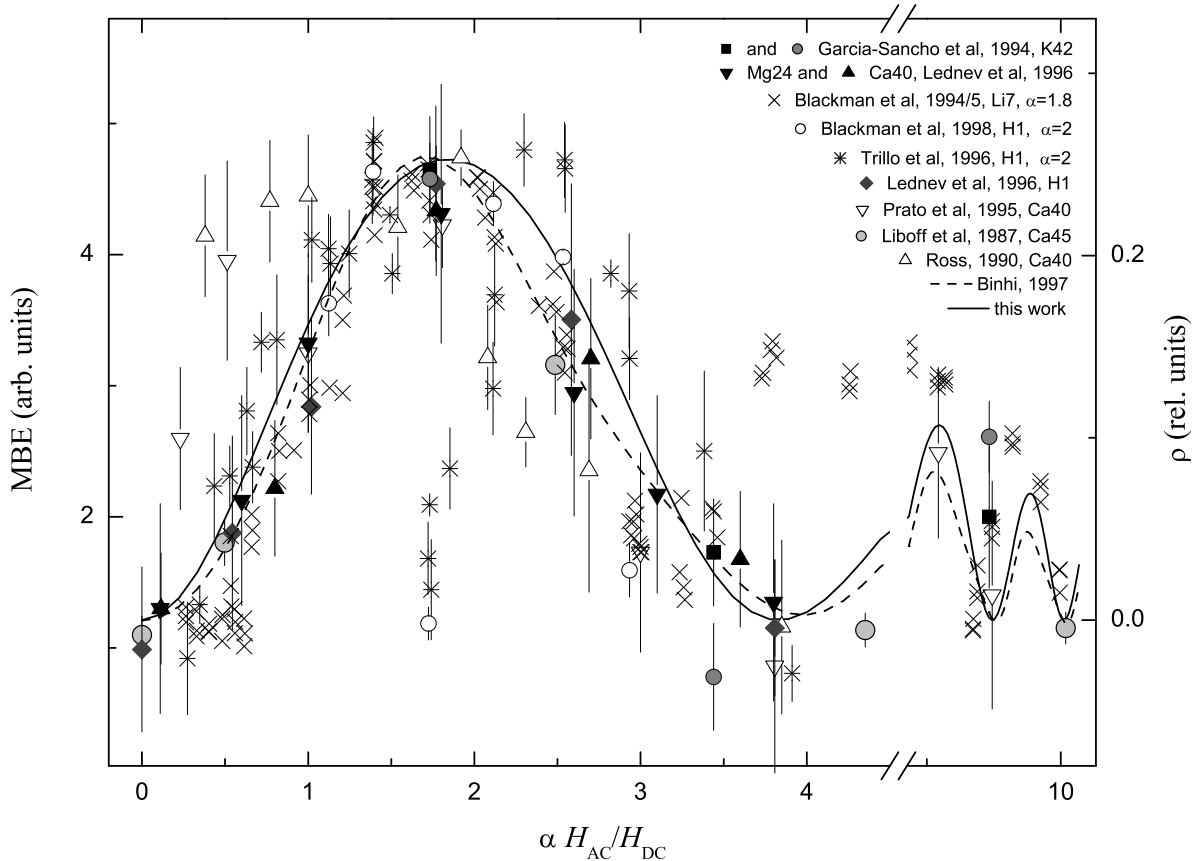,width=0.99\linewidth}
\caption{\label{amplitud}
Experimental evidence
\cite{liboff87b,ross90,blackman94,garcia94,blackman95b,prato95,
lednev96c-e,lednev96b-e,trillo96,blackman98}
for MBEs in a uniaxial MF (\cite{ross90} noted a weak perpendicular
component of a DC MF). Theoretical amplitude spectra: a dash line was derived
for fixed ion-protein complexes (factor $\alpha=1$ was not shown) and also
for rotating ion-protein complexes, see details in \cite{binhi00a}.
Solid line represents the function (\ref{finalf}) derived for gyro interference.}
\end{figure*}

Amplitude `windows', see Fig.\,\ref{amplitud}, specify nonlinearity of the transduction
mechanisms involved in magnetobiological effects. This is confirmed more by
the fact that magnetic noise simultaneously superimposed on a regular
magnetic signal suppresses biological effect of that signal
\cite{mullins93,lin95,raskmark96,litovitz97}.

A nonlinear mechanism based on quantum interference has been developed in
\cite{binhi97} to explain unusual ELF MF frequency and amplitude dependencies
of magnetobiological effects (MBEs). The mechanism elaborates the
interference of ions bound within proteins. According to this mechanism,
superposition of the ion states forms a non-uniform pattern of the
probability density of ion. This pattern consists of a row of more or less
dense segments occurring due to the interference between quantum states of
ions in a protein binding cavity. In a DC MF the pattern rotates with the
cyclotron frequency. Exposure to a time-varying MF of specific parameters
retards the rotation of the pattern and facilitates escape of the ion from
the cavity. This escape might influence equilibrium of biochemical reactions
to ultimately result in a biological effect.

Biologically effective parameters of AC-DC magnetic fields depend on the
charge-to-mass ratio of the ion in question. The closed formula is derived
for `magnetic' part {\sf P} of ion-protein dissociation probability.
Predictions based on this formula reveal good agreement with experimental
results involving calcium, magnesium, potassium, hydrogen and other ions of
as molecular targets for MF. The theory describes multipeak frequency and
amplitude spectra of MBEs involving ions of Ca$^{2+}$, Mg$^{2+}$, and H$^+$
as molecular targets for AC-DC MFs \cite{binhi97}.

The interference mechanism is surprisingly effective in retrospectively
predicting results of existing experiments conducted under the following
defined MF conditions: parallel AC-DC and pulsed MFs \cite{binhi97,binhi98b},
`null' and static MFs \cite{binhi01a}, and various MFs with a slow rotation
of a biological system \cite{binhi00a}. As an example, Fig.\,\ref{amplitud}
demonstrates the comparison of experimental data, in parallel AC-DC MFs, on
MBEs involving fixed and rotating proteins, and calculated curve (dash line).
%---------------------- fig1 -----------------------------------------------
%---------------------- fig1 -----------------------------------------------

The good consistency between theoretical calculations and many experiments
indicates that what underlies magnetobiological effects is most likely an
interference phenomenon.

According to the interference mechanism, the relation should be valid $\Gamma^{-1}
\Omega_{\rm c} \gtrsim 1$, where $\Gamma^{-1}$ is the lifetime of ion quantum states
within a bound cavity and $\Omega_{\rm c}$ is a cyclotron frequency of an ion
in the geomagnetic field, usually 10--100\,Hz. The postulate therefore has to
be made that ion quantum states, more exactly their angular modes, live more
than 0.01 s within the cavity. However it is in contradiction with our common
knowledge that such states might live only $10^{-12}$--$10^{-10}$\,s because
of the thermalizing interaction of ion with cavity walls. On the other hand,
the weak AC MF, $\hbar\Omega_{\rm c} \ll k_{_{\rm B}}T$, is commonly believed
to be unable to contribute into thermally driven (bio)chemical reactions
(so-called kT-problem).

To overcome the problem, we note that there is a specific mechanism that
provides relatively large lifetime of the angular modes. Consider a dipole
molecular group that are attached within the cavity to its walls in two
points, i.e. by two covalent bonds, thus forming a group that may rotate
inside the cavity without contact with walls. Such a construction is referred
to as gyroscope. In the case, it is a molecular gyroscope. Of importance is
the fact that thermal oscillations of that covalent bonds, or gyroscope's
supports, make only zero torque about the axis of rotation. This leads to
relatively slow thermalization of a gyroscopic degree of freedom. Relaxation
is mainly due to van der Waals interaction with thermalizing walls. As far as
the interaction potential, the Lennard--Jones potential, decreases as
$r^{-6}$ and walls' inner surface grows as $r^2$, the overall van der Waals
contribution varies approximately as $r^{-4}$. That is, relaxation quickly
diminishes with the cavity size to grow. Computations show almost free
rotations (thermalization time 0.01\,s) of a molecular gyro within the cavity
of 28 angstr\"{o}ms size. This is enough for the ion interference mechanism
to display itself. Probably, such roomy cavities are formed by ensembles of a
few protein globules, between them, or within some enzymes that unfold DNA
double-helix.

\section{Molecular gyroscope}

A long lifetime of angular modes is the sole serious idealization underlying
the mechanism of ion interference. This idealization would be hard to
substantiate with the ion-in-protein-capsule model. One would have to assume
that the ion forms bound states of the polaron type with capsule walls. In
turn, justification of a large lifetime of polaron angular modes would
require new idealizations. A `vicious circle' occurs which one could not
leave without having to substantially change the model itself. Thus, despite
the obvious advantages of the ion-in-capsule model, namely, simplicity and a
high forecasting skill, we have to recognize its limitations and seek for
other solutions.

\begin{figure}[t]
\centering\epsfig{file=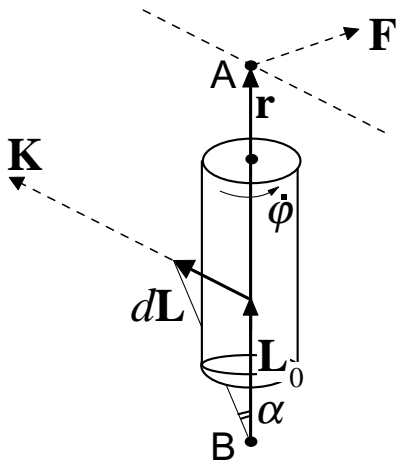,width=0.6\linewidth}
\caption{\label{gyroscop} Forces, moments of forces, and angular momenta in
rotation of a gyro.} \end{figure}

One of them hinges on the use of conservation laws in the dynamics of
rotating solids. Rotation of a solid is described by the equation
\begin{equation} \label{gyr-01} \frac {d {\bf L}} {dt} = {\bf K},
\end{equation} where $\bf L$ is the angular momentum, $\bf K$ is the sum of
torques acting on the solid. Consider for simplicity a symmetric gyro
rotating around one of its main axes of inertia with a force $\bf F$ acting
on its point of support, as shown in Fig.\,\ref{gyroscop}. The moment of this
force about the shown axis is obviously zero. From equation (\ref{gyr-01}) we
have
$$ {\bf L} = {\bf L}_0 + d{\bf L} ,~~~ d{\bf L} = {\bf K}\, dt = {\bf r}
\times {\bf F} dt. $$ Since ${\bf K} \perp {\bf F}$, then $d{\bf L} \perp
{\bf F}$, i.e., the force caused an orthogonal displacement of the axis of
rotation. Also, the vector $\bf r$ is directed along the axis of rotation,
therefore the vector $d{\bf L}$ is also orthogonal with ${\bf L}_0$.
%---------------------- fig2 -----------------------------------------------
%---------------------- fig2 -----------------------------------------------

Thus, a continuously acting force $\bf F$ causes a forced precession of the
gyro about the direction $\bf F$ with an angular velocity defined by the
angle through which the gyro axis of rotation deviates per unit time, viz.,
$$ \Omega_{\rm precession} = \frac {d{\bf L} /{\bf L}_0} {dt} = \frac K {L_0}
= \frac {rF}{L_0}. $$ The length of vector $\bf r$ is defined by the gyro
locking conditions. If point B is fixed, then the origin of $\bf r$ coincides
with B. If point B is free, then the origin of $\bf r$ is on line AB and
depends on the gyro parameters. For estimation, it is important that $r$ has
the order of magnitude of gyro length.

Let the gyro be a model of a rigid molecule free to move and constrained by
the thermal oscillations of one of the point of support (e.g. A) alone. We
estimate the mean gyro axis deviation angle for a random force $\bf F$
causing chaotic oscillations of its point of support. It should be noted that
the gyro gravity energy $\sim MgR$ is many orders of magnitude below its
kinetic energy $\sim L^2/2I$ and the effects of gravity may be neglected. In
the last formulas, $M, R$, and $I$ are the gyro mass, size, and moment of
inertia, and $g$ is the acceleration due to gravity.

The energy of natural gyro rotation is $\varepsilon_0 = L_0^2/2I$. The gyro
energy including chaotic rotations is $\varepsilon_0 + k_{_{\rm B}}T$. On the
other hand, the mean energy with allowance for orthogonality of ${\bf L}_0$
and $d{\bf L}$ is
\begin{eqnarray} \nonumber \langle \frac1{2I} ( {\bf L}_0 +
d{\bf L} )^2 \rangle = \frac1{2I} \left\{ L_0^2 + 2 \langle {\bf L}_0 d{\bf
L} \rangle + \langle d^2 {\bf L} \rangle \right\}\\ = \varepsilon_0 + \frac
{\langle d^2 L \rangle } {2I} , \end{eqnarray}
where brackets mean averaging over the ensemble. Then ${\langle d^2 L \rangle
} /{2I} \sim k_{_{\rm B}}T$. Denoting the average deviation angle by $\alpha
= \sqrt{\langle d^2 L \rangle } /L_0$ yields $\alpha^2 \sim 2Ik_{_{\rm B}}T /
L_0^2$. The smaller $L_0$ the larger the random deviations of a molecule
caused by thermal perturbations of its support. Such a support is the
covalent bond with the body of protein molecule. Low bound estimates of $L_0$
follow from the Heisenberg uncertainty principle which, for a complementary
pair of noncommuting operators of angular variable $\varphi$ and angular
momentum ${\mathcal L} \sim d\,/d\varphi$, can be written as:
$$ \triangle L \, \triangle \varphi \sim
\hbar/2 . $$ Since $ \triangle\varphi \sim \pi $, then $\triangle L \sim
\hbar/2\pi $; thus the angular momentum cannot be smaller than its
uncertainty, i.e., $L_0 \sim \hbar/2\pi$. Finally, we have
$$ \alpha^2 \sim 8\pi^2 \frac {Ik_{_{\rm B}}T} { \hbar^2}. $$ As can be seen
deviations increase with the size of molecule; however, even for small
molecules, the estimate of deviation is unrealistically large. It implies
that, in lower rotation states, molecules will `lay aside' in response to
perturbation of their support and, consequently, the angular momentum will
not be conserved. It should be noted that we are interested only in angular
states with small quantum numbers. Otherwise the interference patterns to be
discussed below become fine grained and are unlikely to be reflected in
measured properties.

Thus, in order to be immune to thermal displacements of supports,
the gyro has to have its second support also fixed in the protein
matrix. The configuration of a rotating solid with supports fixed
in the rim is one of the types of a {\em gyroscope}, i.e., a
device to measure angular displacements and velocities. What we
consider is essentially a molecular gyro: a relatively large
molecular group is placed in a protein cavity and its two edges
form covalent bonds (supports) with the cavity walls. It is
important to note that thermal oscillations of the supports
produce only zero moments of forces about the natural group
rotation axis. Therefore, the gyroscopic degree of freedom
$\varphi$ is not thermalized by the supports' oscillations. This
does not imply that the energy of the gyro does not dissipate.
Radiation damping or Lorentz friction force is neglected, because
of its infinitesimal value. Below we examine at first the
interference of the molecular gyro and then the damping due to wan
der Waals forces.
%---------------------- fig3 -----------------------------------------------
\begin{figure}[th]
\centering\epsfig{file=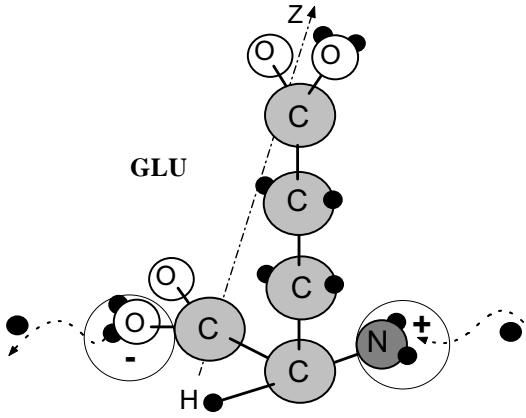,width=0.8\linewidth}
\caption{\label{glu} An amino glutaric acid molecule with
potentially ionizing groups. The $z$ axis is the main axis of
inertia. Rotation of charges distributed over the molecule in a
magnetic field leads to interference of its quantum angular
states.} \end{figure}
%---------------------- fig3 -----------------------------------------------

\section{Interference of the molecular gyroscope}

Rotations of large molecules is much slower a process than electron and
oscillatory processes. Therefore, we think of the rotating molecular group as
a rigid system of charged point masses --- atoms and molecules with partially
polarized chemical bonds. To illustrate, we point to molecules of amino acids
which could be built into rather spacious protein cavities forming chemical
bonds at extreme ends of the molecule, thus forming a molecular gyroscope.
Amino acids are links of polymeric protein macromolecules and also occur in a
bioplasm as free monomers. The general formula of amino acids is well known:
\begin{eqnarray*} & {\rm R}& \\ & | & \\ & ~~~~~ {\rm H}_2 {\rm N}^+ - {\rm
C} {\rm H} - {\rm C} {\rm O} {\rm H} {\rm O}^-& ~, \end{eqnarray*} where R is
a radical which differs one molecule from another. Polarities of the groups
are shown in a water solution. By way of example, the radical of amino
glutaric acid consists of three links $- {\rm C} {\rm H}_2- {\rm C} {\rm
H}_2- {\rm C} {\rm O} {\rm O} {\rm H}$, as shown in Fig.\,\ref{glu}. Fixed
on either side of a cavity, such a molecule, treated as a dynamic unit, has
one degree of freedom --- a polar angle $\varphi$, which simplifies analysis
of its behavior in a magnetic field.

For small velocities, the Lagrange function of one charge particle has the
form
\begin{equation} \label{gyr-1} {\sf L} =\frac{Mv^2}{2}+\frac{q}{c}{\bf
A} {\bf v} - qA_0 , \end{equation} where $\bf v$ is the particle velocity,
and $q$ is a charge. Let the magnetic field $ {\bf H} =(0,~0,~H)$ be directed
along the $z$ axis, and the particle be bounded by a holonomic constraint
causing its circumferential motion in the $xy$ plane. In spherical
coordinates, the constrains can be written in the form
\begin{equation} \label{gyr-2} r=R= \mbox{\rm const.} ,~~~ \theta = {\pi}/2 .
\end{equation} We choose the vector \glossary{vector potential} potential in
the form
\begin{equation} \label{gyr-3} {\bf A}= \left( -\frac12 Hy,~\frac12 Hx,~ 0
\right) . \end{equation} With allowance for constraints (\ref {gyr-2}), the
velocity of a particle in spherical coordinates will be $v=R\dot{\varphi }$,
and the velocity vector in Cartesian coordinates is
\begin{equation} \label{gyr-4} {\bf v}= \left( -R\dot{\varphi }\sin( \varphi
),~ R\dot{\varphi }\cos( \varphi ), ~0 \right). \end{equation} Substituting
this expression in equation (\ref {gyr-1}), we obtain the Lagrange function
in spherical coordinates
\begin{equation} \label{gyr-5} {\sf L} =\frac{MR^2 \dot{\varphi
}^2}{2}+ \frac{qH}{2c}R^2 \dot{\varphi } -qA_0 . \end{equation} Now, the
generalized momentum is $l= \partial{\sf L} / \partial \dot{\varphi }$, and the
Hamilton function ${\sf H} = l\dot{\varphi } -{\sf L} $ is equal to
\begin{equation} \label{gyr-7} {\sf H} = \frac 1{2MR^2}\left(l-
\frac{qH}{2c}R^2 \right)^2 +qA_0~. \end{equation} In the absence of
electromagnetic field ${\sf H} = {l^2}/{2MR^2}$, and it is obvious that $l$
is the angular momentum of the particle. The Hamiltonian operator repeats
(\ref{gyr-7}) with the difference that here $l$ is the angular momentum
operator ${\mathcal L} = -i \hbar{\partial}/{\partial\varphi }$.

Let now a few particles rotate and, in a spherical system of coordinates, the
constraints for particle $i$ be
$$ r_i = \mbox{\rm const.} ~,~~~ \theta_i = \mbox{\rm const.} $$ Then, for a
system of particles in a uniaxial magnetic field, the Lagrange function can
be written following the derivation of formula (\ref{gyr-5}) as
\begin{equation} \label{4-6-1} {\sf L} = \frac{I }2
\dot{\varphi}^2 + \frac{ H Q}{2c} \dot{\varphi} - \sum_i q_i
A_0(r_i,\theta_i,\varphi_i)~, \end{equation} where
\begin{equation} \label{4-6-2} I= \sum_i M_i r_i^2 \sin^2( \theta_i),~~~ Q=
\sum_i q_i r_i^2 \sin^2( \theta_i) \end{equation} is the moment of inertia,
and `charge moment of inertia' of the system about the axis of rotation. As
can be seen, the Lagrange function of the system follows from the Lagrange
function (\ref{gyr-5}) after formal replacement of $MR^2$ with $I$, $qR^2$
with $Q$, and $qA_0$ with the respective sum. Therefore, the Hamiltonian of
the system immediately follows from equation (\ref{gyr-7}) after similar
substitutions
$$ {\mathcal H} = \frac 1{2I}\left( {\mathcal L} - \frac{QH}{2c} \right)^2 +
\sum_i q_i A_0(r_i,\theta_i,\varphi_i) . $$ We assume further that the
electric field is absent, i.e., let $A_0 =0$:
$$ {\mathcal H} = \frac 1{2I}\left( {\mathcal L} - \frac{QH}{2c} \right)^2.
$$ In addition to ${\mathcal L}^2/2I$ we find here two more operators. There
are certain grounds to neglect the term proportional to squared $H$. From the
ratios of coefficients at the terms quadratic and linear in $H$ we obtain
$QH/{4c \hbar} \sim 10^{-7}$, where, for estimation purposes, we let $Q\sim
eR^2$, $R\sim 10^{-7}$\,cm, $H\sim 1$\,G. Dropping this term we write the
Hamiltonian in a convenient form
\begin{equation} \label{4-6-4} {\mathcal H} = \frac{{\mathcal L}^2}{2I} -
\omega (t){\mathcal L} ~,~~~ \omega (t) \equiv \frac{QH}{2Ic} ~.
\end{equation} The eigenfunctions and energies of the time-independent part
of Hamiltonian (\ref{4-6-4}) are
$$ |m\rangle =\frac1{\sqrt {2\pi}} \exp(im\varphi),~~m=0,\pm1,\,...,~~~~
\varepsilon_m =\frac { \hbar^2} {2I}m^2 ~. $$
We now consider the ensemble of gyros that features a density operator
$\sigma$ obeying the Liouville equation
\begin{equation} \label{liou} i\hbar \dot{\sigma} = {\cal H}\sigma - \sigma
{\cal H} ~,~~~ \sigma = \sum_{\alpha} w_{(\alpha)} \sigma^{(\alpha)} ~.
\end{equation}
Some physical quantities, like the intensity of a spontaneous emission or
the radiation reemitted by an ensemble, are known to linearly depend on the
density matrix of the ensemble
\[ \sigma_{mm'} = \sum_{\alpha} w_{(\alpha)} \sigma^{(\alpha)}_{mm'} ~. \]
The probability of biochemical reaction that we examine here is not a
quantity of that sort. The reaction probability does not directly depend on
the density matrix of the ensemble. It is rather the probability of the
reaction of a gyro averaged over that ensemble. Therefore, at first we will
find the density matrix $ \sigma^{(\alpha)}_{mm'}$ of the $\alpha$th gyro,
then the reaction probability of that gyro that non-linearly depends on
$\sigma^{(\alpha)}_{mm'}$, and at last we will average the result over the
gyro ensemble.

Let the ensemble consist of gyros that appear with a constant rate at random
moments of time. We assume the new gyros appear in a quantum state that is a
superposition of the states close to the ground one, i.e.,
\[ \sigma^{(\alpha)}_{mm'}(0) = \left\{ \begin{array}{rl} {\rm const} ~, &
~~ m,m'\sim 1 \\ 0  ~, & ~~ m,m' \nsim 1 \end{array} \right. ~. \]
In the process of thermalization, the levels turn out to be populated with
the energies up to $\varepsilon _m \sim k_{_{\rm B}}T$, i.e., with numbers up
to $m \sim \frac1{\hbar} \sqrt{Ik_{_{\rm B}}T} \sim 10^3$ for gyros with the
inertia moments of the order of $I \sim 10 ^{35}$\,{g$\cdot$cm$^2$}. However,
we are interested in the dynamics of the lowest states, that only could
result in observable effects.

In the representation of the eigenfunctions of ${\cal H}_0$ the density
matrix equation may be written from (\ref{4-6-4}) and (\ref{liou}) as follows
\begin{equation}\label{dotsigma}
\dot{\sigma} _{mm'} = -(\Gamma _{mm'} + i \omega _{mm'} ) \sigma _{mm'} -
\frac i{\hbar} \sum _l ( {\cal V} _{ml} \sigma _{lm'} - \sigma _{ml} {\cal V}
_{lm'} ) ~, \end{equation}
where
\[ \omega _{mm'} = \frac {\hbar} {2I} (m^2 - m'^2) ~,~~~{\cal V} _{ml} =
-\hbar \omega(t) m \delta _{ml} ~. \]
Phenomenological relaxation of the density matrix elements is taken into
account, through the damping constants $\Gamma_{mm'}$. Because of the
relaxation the elements $\sigma _{mm'}$ of the lowest modes decrease while
those of upper modes increase. As far as the stationery dynamics of a separate
gyro is out of interest, we don't allow for the pumping upper modes, i.e.,
population redistribution into the states with large numbers $m$.

Substitution of the above relations in (\ref{dotsigma}) gives rise to the
equation
\[ \dot{\sigma} = -\Gamma \sigma + i \sigma \left[ (m-m') \omega (t) -
\omega \right] ~, \]
where indices $m, m'$ are temporarily omitted for convenience. Along with
notation
\[ g(t) \equiv -\Gamma +i f ~,~~~ f \equiv (m-m') \omega (t) - \omega \]
the equation takes the straightforward form $\dot{\sigma} =g(t) \sigma $.
In the solution of that equation $\sigma = C \exp \left( \int g(t) \, dt
\right)$, the constant $C$ follows starting conditions.

Let the MF possesses both DC and AC parts, then
\[ \omega(t) = \omega_{\rm g} (1+h' \cos\Omega t ) ~, ~~~ \omega_{\rm g}
\equiv \frac {QH_{\rm DC}} {2Ic} ~,~~~ h' \equiv \frac {H_{\rm AC}}
{H_{\rm DC}} ~. \]
Now we separate constant and alternating parts in $g(t)$:
\begin{eqnarray}
g(t) = -x + i z \Omega \cos \Omega t ~,~~~ x \equiv \Gamma +
i\omega - i (m-m') \omega_{\rm g} ~,\nonumber\\
z \equiv (m -m') \omega_{\rm g} \frac {h'} {\Omega} = (m-m') \frac
{h'} {\Omega'} ~, ~~~\Omega' \equiv \frac {\Omega} {\omega_{\rm g}
} ~. \nonumber
\end{eqnarray}
The integral equals
\[ \int g(t)\, dt = \int (-x + iz \Omega \cos \Omega t ) dt = -xt + iz \sin
\Omega t ~, \]
hence
\begin{eqnarray}
 \sigma = \sigma(0) e^{\int g(t) \, dt} = \sigma(0) e^{-xt}
e^{iz \sin \Omega t} \nonumber\\
 =\sigma(0) e^{-xt} \sum_n {\rm J}
_n (z) e^{in \Omega t} ~. \nonumber
\end{eqnarray}
Restoring indeces $m,m'$, we arrive at the equation
\begin{eqnarray}
 \sigma _{mm'} = \sigma_{mm'} (0) e^{- [ \Gamma _{mm'} + i\omega
_{mm'} -i (m-m') \omega_{\rm g} ] t }\nonumber \\
\times \sum_n {\rm J} _n (z_{mm'} ) e^{in \Omega t} ~. \nonumber
\end{eqnarray}
Further, all the damping constants are assumed to equal $\Gamma$. With the
notation
\[ \beta \equiv \Gamma + i \omega_{mm'} -i (m-m') \omega_{\rm g} -in \Omega
~, \]
we rewrite the last equation in the form
\[ \sigma _{mm'} = \sigma_{mm'} (0) \sum_n {\rm J} _n (z_{mm'} ) e^{- \beta
t}  \]
that will be used later.

Now we consider the probability density of a gyro to take an angular
position $\varphi$, which is the only favorable position of the rotating
group of the gyro to react with the active site on the wall
\begin{eqnarray}
p(t) = \Psi^* (t,\varphi ) \Psi (t,\varphi ) = \frac 1{2\pi}
\sum_m c_m^* (t) e^{-im\varphi} \sum_{m'} c_{m'} (t) e^{im'
\varphi}\nonumber\\
 = \frac 1{2\pi} \sum_{mm'} \sigma_{mm'}
e^{-i(m-m')\varphi} ~, \nonumber
\end{eqnarray}
that is,
\[ p(t) = \frac1{2\pi} \sum_{mm'n} \sigma_{mm'} (0) e^{-i(m-m') \varphi}
e^{-\beta t} {\rm J}_n (z_{mm'}) ~. \]
It is expedient to perform a sliding averaging in order to smooth out the
relatively fast oscillations: they do not affect the active site that
features character time constant $\tau$, i.e.,
\[ p_{\tau} (t) = \frac 1{2\tau} \int _{t-\tau} ^{t +\tau} p(t') \, dt' ~. \]
Virtually, the factor $\exp (-\beta t)$ should be averaged:
\[ \left( e^{-\beta t} \right) _{\tau} =  \frac {\sinh (\beta \tau)} {\beta
\tau} e^{-\beta t} ~, \]
therefore
\begin{widetext}
\begin{equation}
\label{ptaut} p_{\tau}(t) = \frac1{2\pi} \sum_{mm'n}
\sigma_{mm'} (0) \frac {\sinh (\beta \tau)} {\beta \tau}  e^{-i(m-m')\varphi}
e^{-\beta t} {\rm J}_n (z_{mm'}) ~.
\end{equation}
\end{widetext}
Then, as in the ion interference model, we assume the reaction probability of
a side group of the rotating molecule with the protein active site to be a
non-linear function of the probability density (\ref{ptaut}). In the absence
of whatever information on that function, it makes sense to consider
quadratic non-linearity, since the linear term makes no contribution to that
probability, see details in \cite{binhi97}. To find the reaction probability
we will square (\ref{ptaut}) and take the average over the gyro ensemble.

In the product $p_{\tau} (t)p_{\tau} (t)$ there are (i) complex conjugate
terms, i.e., pairs with indeces $n,m,m'$ and $-n,m',m$, which apparently do
not oscillate, and (ii) fast-oscillating terms that we omit in view of the
subsequent averaging. Omitting also immaterial numerical coefficient, we
write
\[ p_{\tau} ^2(t) \simeq  e^{-2\Gamma t} \sum _{mm'n} | \sigma_{mm'} (0) |^2
\left| \frac {\sinh (\beta \tau )} {\beta \tau} \right| ^2 {\rm J}_n ^2 (z_{mm'}
) ~. \]
In this expression, the multiplier
\[ S \equiv \sum _{mm'n} | \sigma_{mm'} (0) |^2 \left| \frac {\sinh (\beta
\tau )} {\beta \tau} \right| ^2 {\rm J}_n ^2 (z_{mm'} ) \]
contains the magnetic field dependence.

Let a gyro appear in a moment of time $t'$, then the reaction probability at
time $t$ equals
\[ u(t,t') = \left\{ \begin{array}{rl} S e^{-2\Gamma (t-t')}~, & ~~~ t\geq t'
\\ 0 ~, & ~~~ t < t' ~. \end{array} \right.  \]
Assuming the moments of time $t'$ to be distributed over the gyro ensemble
in the interval $(-\theta,\theta)$ with a uniform density $w$ (instead of a
discrete distribution for $w_{(\alpha)}$ in (\ref{liou})), we find
the mean probability $\sf P$ by proper integrating over the parameter $t'$:
\[ {\sf P} = \lim _{\theta \rightarrow \infty} w \int _{-\theta} ^{\theta}
u(t,t')\, dt' = \frac {w S} {2\Gamma} ~. \]
To link this value to an observable, e.g., a concentration of the reaction
products, we write the kinetic equation for the number $N$ of gyros per unit
of tissue volume
\[ \dot{N} = w - {\sf P} N  \]
that gives $N= w/{\sf P} = 2\Gamma/S$ in stationery conditions. Let $S_0$ and
$N_0$ stand for corresponding quantities in the absence of an AC MF, i.e., at
$h'=0$. We would like to know the relative change $\rho$ of the concentration of the
reaction products under the AC MF influence. This is the relative number of
gyros entering the reaction, i.e.,
\begin{equation}\label{xdef} \rho \equiv \frac {N_0 -N} {N_0} = 1 - \frac {S_0}
{S} ~. \end{equation}
We now estimate values of $S$ and $\rho$. The following notation will be used:
\[ \beta\tau \equiv \eta +i \xi ~,~~~ \eta \equiv \Gamma \tau ~,~~~ \xi
\equiv [\omega_{mm'} - (m-m') \omega _{\rm g} - n\Omega ] \tau ~. \]
Then the expression for $S$ takes the form
\begin{equation}\label{sfin}  S = \sum _{mm'n} | \sigma_{mm'} (0) |^2 \frac
{\sinh ^2 \eta + \sin ^2 \xi} {\eta ^2 + \xi ^2} {\rm J}_n ^2
\left[ (m-m' ) \frac {h'} {\Omega'} \right]. \end{equation}
Since $\eta$ is a constant, the frequency spectrum is defined mainly by the
equation $\xi =0$, i.e.,
\[ \omega_{mm'} -\omega_{\rm g} (m-m') -n\Omega  =0 ~. \]
For arbitrary small $m$, $m'$ frequencies $\omega_{mm'}$ fall into the
microwave range. The effects of low-frequency MFs are defined by the
interference of the levels $m'=-m$, when $\omega_{mm'}=0$. Then
\[ \omega_{\rm g} (m-m') + n\Omega  =0 ~, \]
from which we find
\begin{equation} \label{spectrum} \Omega' = \frac {2m}n ~. \end{equation}
The series over $n$ in (\ref{sfin}) converges quickly, therefore the terms
with $n=1$ mainly contribute to the reaction probability. So, at frequencies
where the probability gains maxima ($\Omega'=2m$) contributions of those
terms equal
\[ S_m \equiv |\sigma_{m,-m}(0)|^2 \frac {\sinh ^2 (\Gamma\tau) } {\Gamma^2
\tau^2 } \, {\rm J}_1 ^2 (h') ~. \]
Contributions of the terms with $n=2$
\[ |\sigma_{m,-m}(0)|^2 \frac {\sinh ^2 (\Gamma\tau) + \sin^2 (6 m \omega
_{\rm g} \tau )} {\tau^2 (\Gamma^2 + 36 m^2 \omega_{\rm g} ^2) } \,
{\rm J}_2 ^2 (2h') ~, \]
obviously, are more than order of value smaller, in the case of $\omega_{\rm
g} \gtrsim \Gamma$, i.e., when it makes sense to examine the interference in
general. Thus, in order to make approximate assessments we omit the terms
with $n > 1$. Then, for the same reason, for the ground state $m=0$ only
contributions of the terms with $n=0$ are essential. It is those terms that
make the contribution independently of an AC MF:
\[ S_0 = |\sigma_{00}(0)|^2 \frac {\sinh ^2 (\Gamma\tau) } {\Gamma^2 \tau^2 }
~. \]
As well, at a fixed frequency $\Omega' = 2m^*$ only terms with $m=-m'=m^*$
are essential in their contribution. Now the relative change of the
concentration of the reaction products is easy to estimate at the MF
frequency, e.g., $\Omega' =2m$. Making note of ${\rm J}_{-1}^2(h') = {\rm
J}_1^2(h')$ and allowing for $S= S_0 + S_m$ in this case, from (\ref{xdef})
we arrive at
\[ \rho = 1- \left[ 1+ 2 \frac {\sigma_{-m,m}^2 (0) } {\sigma_{00}^2 (0)}
{\rm J}_1^2 (h') \right]^{-1} ~. \]
As is seen, the magnitude of the magnetic effect depends on the ratio of
the density matrix elements at the initial moment of time just after a gyro
appears. For example, if the ground state and the state $m$ (out of Zeeman's
splitting)  equipopulated at $t=+0$, then
\begin{equation}\label{finalf}  \rho = 1- \frac1 { 1+ {\rm J}_1^2 (h') } ~.
\end{equation} This function is shown in the Fig.\,\ref{amplitud}, solid line. We
conclude that the positions of the maxima of the amplitude spectrum of the
magnetic effect do not depend (and the relative magnitude of the effect do)
on the distribution of the initial populations of the gyro levels.

The spectrum (\ref{spectrum}) determines only possible locations of extrema.
A real form of the spectrum depends on the initial conditions for the density
matrix, i.e., on the populations of levels of different rotational quantum
number $m$.

It is instructive to note that the molecule need not have a dipole moment
$\sum_i q_i {\bf r}_i$ for the magnetic effect to appear. Rather, it is important
that the `charge moment of inertia' $Q$ (\ref{4-6-2}) be other than zero. This
can be the case in the absence of dipole moment, e.g. for ionic rather than
zwitterionic form of the molecule.

The main properties of the gyro interference are identical with those of the
ion interference, namely, (i) multiple peaks in the amplitude and frequency
spectra, (ii) dependence of the positions of frequency peaks on the DC MF
intensity, and (iii) independence of the positions of amplitude maxima on the
AC MF frequency.

We note that the interference of a molecular gyro has some features that
differ it from the ion interference. Firstly, the peak frequencies are
defined with respect to the gyral frequency $\omega_{\rm g}$ --- a rotation
equivalent of cyclotron frequency. Peak frequencies depend on the
distribution of electric charges over the molecule and may deviate from
harmonics and subharmonics of the cyclotron frequency. Secondly, the gyro
rotation axis is fixed with respect to the shell, which introduces, in the
general case, one more averaging parameter in the model. However, these
features are not of principal significance. The specific properties of the
interference can always be calculated for any configuration of magnetic and
electric fields, for rotation of biological systems and macromolecules
involved, etc.

There is the crucial feature of the gyro interference: molecular gyros are relatively
immune to thermal shaking and may be effective biophysical targets for external MFs.

As is seen from (\ref{sfin}) the absolute magnitude of the magnetic effect,
where the latter is maximized by the MF parameters, depends mainly on the
value $\eta = \Gamma \tau$, which should be minimized for greater effects.
The protein reaction time $\tau$ and the MF frequency $\Omega$ have to
fulfill the relation $\Omega\tau \gtrsim 1$ in order to manifest an interference.
This and the properties of the function $\sinh^2 \eta /\eta^2$ lead to the
condition of observability
\begin{equation}
\Gamma^{-1} \gtrsim \Omega^{-1} \sim 0.01 \,\mbox{\rm s}
\end{equation}
for the ELF range. The following section examines if the condition is real.

\section{Estimating relaxation time from molecular dynamics}

Computer simulation of molecular gyro behavior indicates that, for relaxation
times of order 0.01\,s, the size of cavity should be below 30\,{\AA}.

\begin{figure}[t]
\centering\epsfig{file=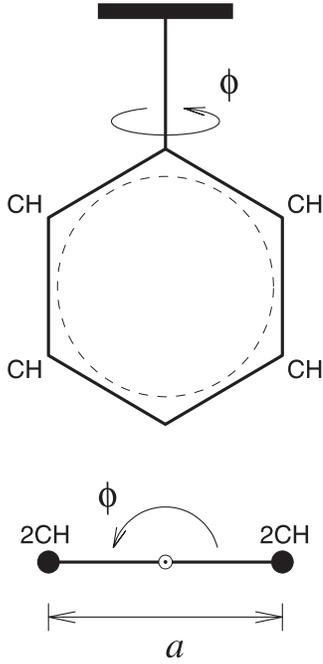,width=0.5\linewidth}
\caption{\label{benzol} Schematic representation of a molecular
gyroscope C$_{\alpha}$C$_6$H$_5$ reduced to a model of a
two-particle rotor of diameter $a$.} \end{figure}

We consider the amino acid residue Phenilalanin, (Phe) C$_\alpha$C$_6$H$_5$,
as a gyro and look at the revolution of its benzene ring C$_6$H$_5$ about the
valence bond C$_{\alpha}$---C$_{\beta}$ --- see Fig.\,\ref{benzol}. This
revolution may be thought of as a rotation in one plane of two rigidly bound
point masses $m=26\,m_{\rm p}$ ($m_{\rm p}$ is the mass of proton), spaced by
$a=2.42$\,{\AA} from one another, about their common center of gravity.

\begin{figure}[t]
\centering\epsfig{file=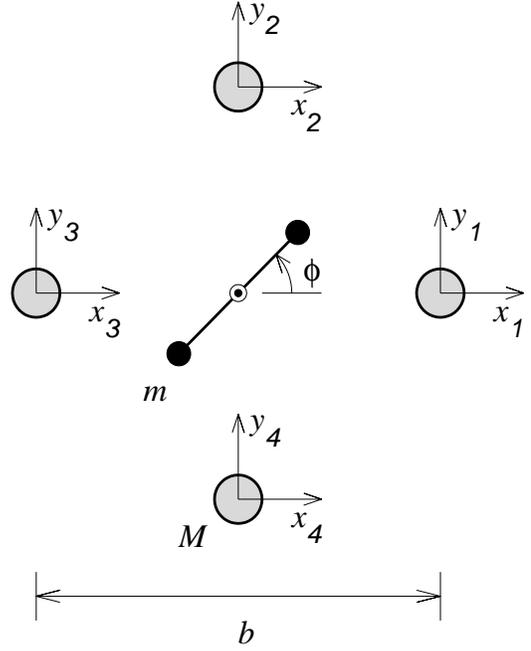,width=0.8\linewidth}
\caption{\label{cavity}\protect A two-dimensional model of a gyro
in a molecular cavity of diameter $b$ formed by four heavy
particles of mass $M$.}
\end{figure}

We model the cavity by four heavy particles of mass $M\ge m$ placed in the
corners of a square (diagonal $b>a$) centered on the gyro axis, as shown in
Fig.\,\ref{cavity}. We assume that these particles oscillate in the gyro
rotation plane $xy$. Each particle moves in the potential well $U(x_i,y_i)$,
where $x_i$, $y_i$ is the deviation of particle $i$ from its equilibrium
state. The Hamilton function for this system has the form
\begin{eqnarray}
{\sf H} &=& \frac12 I\dot{\phi}^2+\sum_{i=1}^4 [ \frac12M (\dot{x}_i^2 +\dot{y}_i^2)
+V(\phi,x_i,y_i)\nonumber \\ &&~~~~~~~~~~~~~~~~~~~~~~~~~~~~ +U(x_i,y_i)],
\label{f1} \end{eqnarray} where $I=\frac12 ma^2$ is the gyro moment of
inertia, and $\phi$ is its revolution angle.
%---------------------- fig4 -----------------------------------------------
%---------------------- fig4 -----------------------------------------------

We take the potential of interaction of particle $i$ with the gyro as the sum
of two Lennard--Jones potentials
$$ V(\phi,x_i,y_i)=\epsilon\{
[(r_0/r_1)^6-1]^2+[(r_0/r_2)^6-1]^2\}, $$ where $r_0$ is the equilibrium arm
between a heavy and a light particle, $r_1$ is the instantaneous distance of
a heavy particle $i$ to the first particle of the gyro, and $r_2$ is the
distance to the second particle. The interaction of carbon atoms in polymeric
macromolecules is commonly described by Lennard--Jones potentials of the form
$$ V_{\rm LJ}(r)=4\epsilon_0 [(\sigma/r)^{12} -(\sigma/r)^6] $$ with
$\sigma=3.8$\,\AA and $\epsilon_0=0.4937$\,kJ/mol \cite{noid91,savin98}.
Recognizing that each particle of the gyro consists of two carbon atoms, we
let $\epsilon=1$\,kJ/mol $\approx 2\epsilon_0$ and $r_0=4.5$\,\AA $\approx
2^{1/6}\sigma$.
%---------------------- fig5 -----------------------------------------------
%---------------------- fig5 -----------------------------------------------

The carrier potential for each heavy particle will be taken in the form
$$ U(x,y)=\frac12K \frac{x^2+y^2} {1-(x^2+y^2)/R_0}, $$ where $K$ is the
rigidity in particle-carrier interaction, and $R_0$ is the maximum possible
deviation radius of a heavy particle. In a protein macromolecule, the
rigidity of atomic displacements is $K=4$\,N/m. We consider two maximum
displacement values: $R_0=1$\,\AA\ and $R_0=\infty$.

Assuming that heavy particles alone are connected with the thermostat, we
obtain the equations of motion in the form
\begin{eqnarray} I\ddot{\phi}&=&-\frac{\partial {\sf H}}{\partial \phi}, \nonumber
\\ M\ddot{x}_i &=& -\frac{\partial {\sf H}}{\partial x_i}-\Gamma_{\rm r} M\dot{x}_i+\xi_i,
\label{f2} \\ M\ddot{y}_i &=& -\frac{\partial {\sf H}}{\partial y_i}-\Gamma_{\rm r}
M\dot{y}_i+\eta_i, \nonumber \\ &~& ~~~~~~~~~ i=1,2,3,4~, \nonumber
\end{eqnarray} where the system's Hamilton function is given by equation
(\ref{f1}); $\xi_i$ and $\eta_i$ are random normally distributed forces
(white noise) describing the interaction of a heavy particle $i$ with the
thermostat, $\Gamma_{\rm r} =1/t_{\rm r}$ is the friction factor, and $t_{\rm r}$ is
the particle velocity relaxation time. The correlation functions of random
forces are
\begin{eqnarray} \langle\xi_i(t_1)\xi_j(t_2)\rangle &=& 2M\Gamma_{\rm r} k_{_{\rm
B}}T\delta_{ij}\delta(t_1-t_2), \nonumber \\ \langle \eta_i(t_1)\eta_j(t_2)
\rangle &=& 2M\Gamma_{\rm r} k_{_{\rm B}}T\delta_{ij}\delta(t_1-t_2), \nonumber \\
\langle\xi_i(t_1)\eta_j(t_2)\rangle &=& 0~. \nonumber \end{eqnarray} Here,
$k_{_{\rm B}}$ is the Boltzmann constant, and $T$ is the thermostat
temperature.

We integrate the equation system (\ref{f2}) by the Runge--Kutta method to the
fourth order of accuracy with a constant integration step $\Delta t$. In this
computation, the delta function $\delta(t)$ is $0$ for $|t| > \Delta t/2$ and
$1/\Delta t$ for $|t| < \Delta t/2$, that is, the integration step
corresponds to the correlation time of random force. Therefore, to use a
system of Langevin equations, we need that $\Delta t \ll t_{\rm r}$. Let the
relaxation time be $t_{\rm r}=0.2$\,ps, and the numerical integration step be
$\Delta t=0.0025$\,ps.
%---------------------- fig6 -----------------------------------------------
\begin{figure}[t]
\centering\epsfig{file=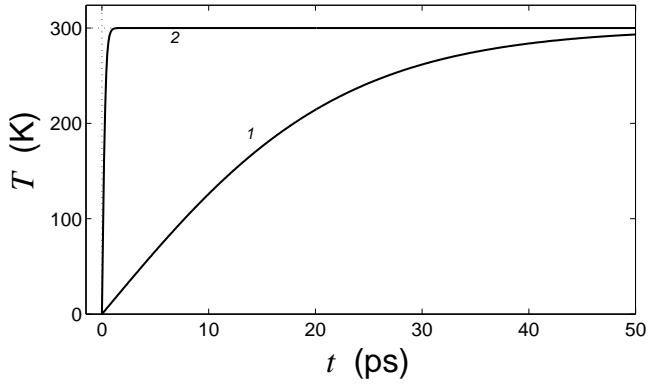,width=1\linewidth}
\caption{\label{timedep} Current mean temperature of the molecular
thermostat $T_1$ (curve 1) and temperatures of its molecular
neighborhood $T_2$ (curve 2) as functions of time. Thermostat
temperature $T=300$\,K, cavity diameter $b=11$\,\AA, $M=m$, and
$R_0=1$\,\AA.} \end{figure}
%---------------------- fig6 -----------------------------------------------

Let in the initial moment of time $t=0$ the system be in the fundamental
state
\begin{eqnarray} \label{f3} \phi(0)=\phi_0,~~x_i(0)=u_i,~~y_i(0)=v_i,
\\ ~~\dot{\phi}(0) = 0,~~\dot{x}_i(0) = 0,~~\dot{y}_i(0) = 0,~~i=1,2,3,4~,
\nonumber \end{eqnarray} where the coordinates of a steady state, $\phi_0$,
$\{u_i,v_i\}_{i=1}^4$, are determined as solutions to the minimization
problem
\begin{equation} {\sf H} \rightarrow \min_{\phi,x_1,...,y_4}:
\dot{\phi}\equiv 0, \dot{x}_1\equiv 0, ..., \dot{y}_4\equiv 0~. \nonumber
\end{equation} Thus, at time zero, the molecular gyro is not thermalized.

Our objective is to estimate the average time of gyro thermalization. It
corresponds to the relaxation time of gyro rotation in a thermalized
molecular system. For this purpose, we numerically integrate the equations of
motion (\ref{f2}) subject to the initial condition (\ref{f3}).

The gyro thermalization at time $t$ is characterized by its current
temperature
$$ T_1(t)=I\langle \dot{\phi}^2(t)\rangle/k_{_{\rm B}}, $$ where
brackets $\langle\cdot\rangle$ imply averaging over independent realizations
of random forces $\xi_i(t)$, $\eta_i(t)$, $i=1,2,3,4$. To obtain the average
value, the system (\ref{f2}) was integrated more than 10000 times.

In turn, the thermalization of the system of heavy particles is characterized
by its current temperature
$$ T_2(t)=\frac{M}{8k_{_{\rm B}}}\sum_{i=1}^4 \langle \dot{x}_i^2(t) +
\dot{y}_i^2(t)\rangle . $$ The time dependence of these temperatures is
presented in Fig.\,\ref{timedep}. At $t=0$, the temperatures are
$T_1(0)=T_2(0)=0$. Further on the time coordinate, they monotonously approach
the thermostat temperature $T=300$\,K.

We will assume that the molecular subsystem is completely thermalized if its
current temperature exceeds $0.99T$. We determine the gyro thermalization
time $t_1$ as a solution of the equation $T_1(t)=0.99T$, and the time of
heavy particle system thermalization $t_2$, as a solution of the equation
$T_2(t)=0.99T$. The gyro is thermalized by interacting with the system of
heavy particles, therefore its thermalization time will depend on the
diameter $b$ of the heavy particle system and will always exceed the heavy
particle system thermalization time $(t_1>t_2)$. Time $t_2$ is almost
independent of $b$ and is dependent only on the relaxation time $t_{\rm r}$:
$t_2\approx 4t_{\rm r}$.

\begin{figure}[t]
\centering\epsfig{file=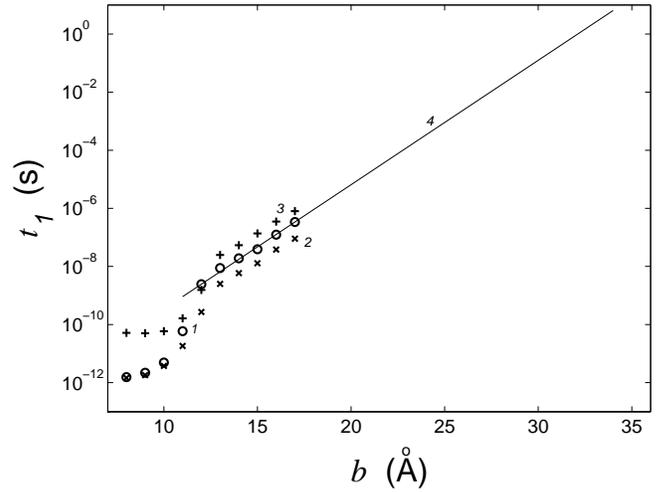,width=1\linewidth}
\caption{\label{thermal}
Gyro thermalization time $t_1$ computed as a function of molecular cavity
diameter $b$ at $M=m$, $R_0=1$\,{\AA} (symbols 1); $M=m$, $R_0=\infty$
(symbols 2); $M=100m$, $R_0=\infty$ (symbols 3), and extrapolation of this
function to large $b$ (curve 4). } \end{figure}

We analyzed the behavior of the system for $R_0=1$\,{\AA}, $\infty$ and
$M=m$, $100m$. The dependence of gyro thermalization time $t_1$ on cavity
diameter $b$ is shown in Fig.\,\ref{thermal}. It is evident that, whatever
the values of $R_0$ and $M$, the thermalization time increases exponentially
with $b$. If we extrapolate this dependence to the range of large $b$, we see
that, at $b=28$--32\,{\AA}, the thermalization time, and hence the gyro
relaxation time $\Gamma^{-1}$, will be of the order of seconds. With this
size of cavity, the molecular gyro will revolve almost freely.
%---------------------- fig7 -----------------------------------------------
%---------------------- fig7 -----------------------------------------------

\section{Conclusion}

The molecular interfering gyroscope is a challenger for solving the
kT-problem as a probable mechanism of magnetobiological effects. Indeed, the
walls of a protein cavity do not interfere with the gyro degree of freedom
directly via short-range chemical bonds. For cavities larger than 30\,{\AA}
in size, the contribution to the relaxation from the van der Waals
electromagnetic forces, induced by wall oscillations, is small. Radiation
damping is negligibly small. Finally, the oscillations of gyro supports
produce a zero moment of forces about the axis of rotation and do not affect
the angular momentum. The gyro degree of freedom is very slow to thermalize,
its dynamic behavior is coherent, which gives rise to slow interference
effects. Of course, whether or not some more or less water-free cavities of
the size of 30\,{\AA} and larger do exist remains an open question, but, what
is essential, ELF magnetic field bioeffects are no longer a paradox.

The role of molecular gyros could probably be played by short sections of
polypeptides and nucleic acids built inside globular proteins or in cavities
between associated globules. In this respect it is interesting to look at
Watson--Crick pairs\glossary{Watson--Crick pairs} of nitrous bases
(adenine--thymine and guanine--cytosine) which bind DNA strands into a double
helix as well as some other hydrogen-bound complexes of nitrous bases. Their
rotations are hampered by steric factors. However, in the realm of activity
of special DNA enzymes, steric constraints may be lifted to allow a
relatively free rotation of molecular complexes. It is not yet clear whether
or not the gyro type of molecular structures exists. They are unlikely to be
detected by X-ray methods since these require crystallization of proteins for
structural analysis. In this state, the rotation would likely be frozen.
Should a rotation be allowed, the mobile groups would not give clear cut
reflections. Some other methods are needed that would work with native forms
of proteins avoiding distortions due to crystallization.

Generally speaking, the fact that the molecular gyro model gives a physically
consistent explanation of MBEs proves indirectly its real grounds. Further
studies should verify whether this conclusion is correct. In any case, today,
the interfering molecular gyroscope is a single available mechanism to give
explanations that would be physically transparent and generally agreeable
with experiments.

%\bibliography{refs}
%-------------------------------------------------------------------------

%---------------------------------------------------------------------------

\newpage

\end{document}